\documentclass[aps,prb,twocolumn,showpacs,floatfix]{revtex4}

\usepackage{graphicx}

\begin{document}

\title{The $t$-$J$ model on a semi-infinite lattice}

\author{A.~Sherman}
\author{N.~Voropajeva}

\affiliation{Institute of Physics, University of Tartu, Riia 142,
 51014 Tartu, Estonia}
\date{\today}

\begin{abstract}
The hole spectral function of the $t$-$J$ model on a two-dimensional semi-infinite lattice is calculated using the spin-wave and noncrossing approximations. In the case of small hole concentration and strong correlations, $t\gg J$, several near-boundary site rows appear to be depleted of holes. The reason for this depletion is a deformation of the magnon cloud, which surrounds the hole, near the boundary. The hole depletion in the boundary region leads to a more complicated spectral function in the boundary row in comparison with its bulk shape.
\end{abstract}


\maketitle

\section{Introduction}
In recent years, an active interest is taken in the electronic properties of heterostructures and surfaces of strongly correlated materials. \cite{dagotto} Looking for new effects and their possible applications a wide variety of systems has been investigated both experimentally and theoretically. Theoretical studies of charge excitations near the crystal boundary have been carried out mainly in the framework of the two- (2D) and three-dimensional (3D) Hubbard model. For this purpose different approximate methods have been used, including the slave boson method, \cite{hasegawa} the perturbation theory \cite{potthoff97} and the dynamical mean-field theory. \cite{potthoff99,ishida} In these works, the case of half-filling was considered, when strong electron correlations cause the antiferromagnetic ordering of the crystal. \cite{hirsch85} However, approximations used in the mentioned works did not take into account
the ordering and the interaction of electrons with respective magnetic excitations. One of the results obtained in Refs.~\onlinecite{hasegawa,potthoff97,potthoff99,ishida} for uniform model parameters is that on the surface layer the quasiparticle weight is smaller than the bulk value. The reason is a reduced surface coordination number which implies a lower kinetic energy and consequently effectively stronger correlation effects at the surface.

It is known \cite{dagotto94} that in the case of strong electron correlations the interaction with magnetic excitations plays an important role in the formation of the low-frequency dispersion of charge carriers. Therefore, peculiarities of these excitations in the near-boundary region may have a significant impact upon the properties of electrons here. The magnetic excitations are described by the quantum Heisenberg model. \cite{dagotto94} The influence of boundaries on its spectrum and observables has been studied in two  \cite{hoglund,metlitski,pardini,voropajeva} and three \cite{sherman09} dimensions. In particular it was shown that absolute values of the nearest-neighbor spin correlations near the boundary exceed the bulk value. In Refs.~\onlinecite{voropajeva,sherman09} this result was related to the peculiar spectrum of the semi-infinite $d$-dimensional antiferromagnet. The spectrum involves $d$-dimensional bulk modes -- standing spin waves -- and a $(d-1)$-dimensional mode of boundary spin waves. These latter excitations eject the bulk excitations from the near-boundary region. Thus the antiferromagnet appears to be divided into two regions with different dominant spin excitations. Charge carriers in the near-boundary region and deep within the crystal appear to be in different spin-excitation environments, that inevitably leads to a dissimilarity in properties of these carriers. Another effect which can contribute to this difference is a smaller number of spin bonds destroyed by charge carriers near the boundary in comparison with the bulk. As will be seen below, this leads to an attraction of the quasiparticles to the boundary.

To answer the question on how the above-mentioned factors influence the distribution of charge carriers near the boundary we consider the 2D $t$-$J$ model, having in mind a Cu-O plane of hole-doped cuprate perovskites. To calculate the hole Green's function in the case of strong correlation, $t\gg J$ ($t$ and $J$ are the nearest neighbor hopping and exchange constants) we apply the spin-wave and noncrossing approximations which were successfully used for unbounded crystals. \cite{marsiglio,martinez} An obtained self-energy equation for the hole Green's function is solved by iterations. Since the translation invariance is violated in the $x$ direction perpendicular to the boundary, the spectral function, apart from the frequency and the $y$ component of the wave vector, depends on the $x$ coordinates of site rows for which the function is considered. With a change of the $x$ coordinates from 0 (the boundary) deep into the crystal the intensity is redistributed in the function and the main maximum is enhanced and shifted to lower frequencies. Such behavior of the maximum indicates that the near-boundary region is depleted of holes at low hole concentrations. The appearance of this depletion is connected with the character of hole excitations. They are spin polarons in which a hole is surrounded by a cloud of magnons. Near the boundary, this cloud is deformed, which leads to an energy loss and to the observed shift of the main maximum to higher energies. Another consequence of the frequency-separated maxima in neighboring rows is a more complicated structure of the boundary spectral function in comparison with its bulk counterpart. The reason is a replica of a stronger maximum from the second row which is also seen in the boundary spectral function.

\section{Main formulas}
Our starting point is the Hubbard Hamiltonian on a square semi-infinite lattice. We consider an idealized boundary which is located along the $y$ crystallographic axis. The variation of the lattice spacing and model parameters near the boundary is neglected. The Hamiltonian reads
\begin{eqnarray}
H_H&=&t\sum_{l_y\delta\sigma}\sum_{l_x\geq 0}a^\dagger_{l_y+\delta,l_x \sigma}a_{l_yl_x\sigma}\nonumber\\
&&+t\sum_{l_y\sigma}\sum_{l_x\geq 0}\left( a^\dagger_{l_y,l_x+1,\sigma}a_{l_yl_x\sigma}+a^\dagger_{l_yl_x\sigma} a_{l_y,l_x+1,\sigma}\right)\nonumber\\
&&+U\sum_{l_y,l_x\geq 0}n_{l_yl_x,+1}n_{l_yl_x,-1}- \mu\sum_{l_y\sigma}\sum_{l_x\geq 0}n_{l_yl_x\sigma},
\label{hubbard}\end{eqnarray}
where $a_{l_yl_x\sigma}$ is the electron annihilation operator, $l_y$ and $l_x$ are the vector components labeling sites of the crystal, which is located at $l_x\geq 0$, $\sigma=\pm 1$ is the spin projection, $\delta=\pm 1$, the lattice spacing is set as the unit of length, $U$ is the Hubbard on-site repulsion, $n_{l_yl_x\sigma}=a^\dagger_{l_yl_x\sigma} a_{l_yl_x\sigma}$, and $\mu$ is the chemical potential. Only the hopping between nearest neighbor sites $t$ is taken into account in Eq.~(\ref{hubbard}).

In the case of strong electron correlations, $U\gg t$, and an electron filling less than half-filling Hamiltonian~(\ref{hubbard}) can be reduced to the Hamiltonian of the $t$-$J$ model using the known unitary transformation \cite{hirsch} $H_{tJ}=e^SH_He^{-S}$ with
\begin{eqnarray*}
S&=&\frac{t}{U}\sum_{l_y\delta\sigma}\sum_{l_x\geq 0}\sigma\left( X^{2,-\sigma}_{l_y+ \delta,l_x}X^{0\sigma}_{l_yl_x}-X^{\sigma 0}_{l_y+\delta,l_x}X^{-\sigma,2}_{l_yl_x}\right)\nonumber\\
&&+\frac{t}{U}\sum_{l_y\sigma}\sum_{l_x\geq 0}\sigma\left( X^{2,-\sigma}_{l_y,l_x+1} X^{0\sigma}_{l_yl_x}-X^{\sigma 0}_{l_y,l_x+1}X^{-\sigma,2}_{l_yl_x}\right.\nonumber\\
&&\quad\quad+\left. X^{2,-\sigma}_{l_yl_x}X^{0\sigma}_{l_y,l_x+1}-X^{\sigma 0}_{l_yl_x}X^{-\sigma,2}_{l_y,l_x+1}\right),
\end{eqnarray*}
where the Hubbard operators \cite{izyumov}
$$
a_{l_yl_x\sigma}=X^{0\sigma}_{l_yl_x}+\sigma X^{-\sigma,2}_{l_yl_x}, \quad a^\dagger_{l_yl_x\sigma}=X^{\sigma 0}_{l_yl_x}+\sigma X^{2,-\sigma}_{l_yl_x}
$$
were introduced. Up to the terms of the second order in $\frac{t}{U}$ the transformed Hamiltonian reads
\begin{eqnarray}
H_{tJ}&=&t\sum_{l_y\delta\sigma}\sum_{l_x\geq 0}X^{\sigma 0}_{l_y+\delta,l_x}X^{0\sigma}_{l_yl_x}\nonumber\\
&&+t\sum_{l_y\sigma}\sum_{l_x\geq 0}\left(X^{\sigma 0}_{l_y, l_x+1}X^{0\sigma}_{l_yl_x}+X^{\sigma 0}_{l_yl_x} X^{0\sigma}_{l_y,l_x+1}\right)\nonumber\\
&&+J\sum_{l_y,l_x\geq 0}\biggl({\bf S}_{l_y+1,l_x}{\bf S}_{l_yl_x}+ {\bf S}_{l_y,l_x+1}{\bf S}_{l_yl_x}\nonumber\\
&&\quad\quad-\frac{1}{4}n_{l_y+1,l_x}n_{l_yl_x}-\frac{1}{4}n_{l_y,l_x+1} n_{l_yl_x}\biggr)\nonumber\\
&&+\mu\sum_{l_y,l_x\geq 0}X^{00}_{l_yl_x},
\label{tJ}\end{eqnarray}
where ${\bf S}_{l_yl_x}$ is the spin-$\frac{1}{2}$ operator, $J=\frac{4t^2}{U}$, and $n_{l_yl_x}=\sum_\sigma n_{l_yl_x\sigma}= 1-X^{00}_{l_yl_x}$ in the considered approximation in which terms containing doubly occupied site states are neglected. In Eq.~(\ref{tJ}), we neglected also terms proportional to $J$ which describe an assistant hole hopping (three-site terms), as it is frequently done in the consideration of the $t$-$J$ model.

Further simplifications of the model Hamiltonian are connected with the spin-wave approximation which in application to the $t$-$J$ model was shown to give results in good agreement with exact diagonalization. \cite{marsiglio,martinez} In the case of low doping and zero temperature the crystal has the long-range antiferromagnetic ordering and the simplest version of the spin-wave approximation can be applied using the following Holstein-Primakoff representation: \cite{tyablikov}
\begin{eqnarray}
S^z_{l_yl_x}&=&e^{i\pi(l_y+l_x)}\left(\frac{1}{2}-b^\dagger_{l_yl_x} b_{l_yl_x}\right),\nonumber\\
S^+_{l_yl_x}&=&P^+_{l_yl_x}\varphi_{l_yl_x}b_{l_yl_x}+P^-_{l_yl_x} b^\dagger_{l_yl_x}\varphi_{l_yl_x},\label{swa}\\
S^-_{l_yl_x}&=&P^-_{l_yl_x}\varphi_{l_yl_x}b_{l_yl_x}+P^+_{l_yl_x} b^\dagger_{l_yl_x}\varphi_{l_yl_x},\nonumber
\end{eqnarray}
where the spin-wave operators $b_{l_yl_x}$ and $b^\dagger_{l_yl_x}$ satisfy the Boson commutation relations and
$$P^\pm_{l_yl_x}=\frac{1}{2}\left(1\pm e^{i\pi(l_y+l_x)}\right), \quad \varphi_{l_yl_x}=\sqrt{1-b^\dagger_{l_yl_x}b_{l_yl_x}}.$$
In the considered antiferromagnetic background the hole creation operator is defined as
$$h^\dagger_{l_yl_x}=\sum_\sigma P^\sigma_{l_yl_x} X^{0\sigma}_{l_y l_x}.$$
Using this definition and Eq.~(\ref{swa}) in Hamiltonian~(\ref{tJ}) and leaving terms up to the second order in the spin-wave operators we get
\begin{eqnarray}
H&=&t\sum_{l_y\delta}\sum_{l_x\geq0}h_{l_y+\delta,l_x}h^\dagger_{l_yl_x} \left(b_{l_yl_x}+b^\dagger_{l_y+\delta,l_x}\right)\nonumber\\
&&+t\sum_{l_y,l_x\geq0}\left[h_{l_y,l_x+1}h^\dagger_{l_yl_x} \left(b_{l_yl_x}+b^\dagger_{l_y,l_x+1}\right)\right.\nonumber\\
&&\quad\quad\quad\left. +h_{l_yl_x}h^\dagger_{l_y,l_x+1} \left(b_{l_y,l_x+1}+ b^\dagger_{l_yl_x}\right)\right]\nonumber\\
&&+H_{AF}-\frac{J}{4}\sum_{l_y,l_x\geq0}\left(\nu_{l_y+1,lx}\nu_{l_ylx} +\nu_{l_y,lx+1}\nu_{l_ylx}\right)\nonumber\\
&&-\frac{J}{2}\sum_{l_y}\nu_{l_y0}+\mu\sum_{l_y,l_x\geq0}\nu_{lylx},
\label{hsw}\end{eqnarray}
where $\nu_{l_yl_x}=h^\dagger_{l_yl_x} h_{l_yl_x}$ and
\begin{eqnarray}
H_{AF}&=&J\sum_{l_y,l_x\geq0}\left[2\left(1-\frac{1}{4}\delta_{l_x0} \right)b^\dagger_{l_yl_x}b_{l_yl_x}\right.\nonumber\\
&&\quad\quad\quad+\frac{1}{2}\left(b_{l_y+1,l_x}b_{l_yl_x}+ b^\dagger_{l_y+1,l_x}b^\dagger_{l_yl_x}\right)\nonumber\\
&&\quad\quad\quad+\left.\frac{1}{2}\left(b_{l_y,l_x+1}b_{l_yl_x}+ b^\dagger_{l_y,l_x+1}b^\dagger_{l_yl_x}\right)\right]
\label{haf}\end{eqnarray}
is the Hamiltonian of the 2D semi-infinite Heisenberg antiferromagnet in the spin-wave approximation. In Eq.~(\ref{hsw}), some constant terms were omitted and the term $\frac{3}{2}J$ was added to the chemical potential.

The next to last term in the right-hand side of Eq.~(\ref{hsw}) describes an attraction of a hole to the boundary. It originates from terms of Hamiltonian~(\ref{tJ}) which contain $z$ components of spins and occupation numbers on neighboring sites. In the antiferromagnetic state, these terms give the energy gain equal to $\frac{J}{2}$ for each nearest-neighbor bond. In the 2D case a hole destroys 4 such bonds deep inside the crystal and 3 bonds on the boundary. Thus, for a hole it is energetically more favorable to reside at the boundary.

Refusing the constraint $l_x\geq0$ and carrying out the Fourier transformation over the space coordinates, Eq.~(\ref{hsw}) is reduced to the spin-wave Hamiltonian on an unbounded lattice, used in Refs.~\onlinecite{marsiglio,martinez} and in a lot of subsequent works.

Considering the case of a low hole doping, in Hamiltonian~(\ref{hsw}) we shall neglect terms containing two hole occupation operators on neighboring sites. Our aim is to calculate the hole Green's function
$$G(k_y\tau l_xl'_x)=-\left\langle{\cal T}h_{k_yl_x}(\tau) h^\dagger_{k_yl_x}\right\rangle,$$
where the angular brackets denote the statistical averaging, ${\cal T}$ is the time-ordering operator that arranges other operators from right to left in ascending order of times $\tau$, $h_{k_yl_x}$ is the Fourier transform of $h_{l_yl_x}$, and $h_{k_yl_x}(\tau)=e^{\tau H}h_{k_yl_x} e^{-\tau H}$. For this calculation, we use the diagram technique with the expansion in powers of $t$, also in full analogy with what was done for the unbounded crystal. \cite{marsiglio,martinez} For this latter case, the self-energy equation was obtained in the noncrossing (Born) approximation in which diagrams with intersecting magnon lines were neglected. It was shown that results obtained in this approximation are in good agreement with data of exact diagonalization. Therefore, we also use this approximation and obtain the following self-energy equation:
\begin{eqnarray}
G(k_ynl_xl'_x)&=&G^{(0)}(nl_xl'_x)
+\sum_{l_{x1},l_{x2}\geq0} G^{(0)}(nl_xl_{x1})\nonumber\\
&&\times\Sigma(k_ynl_{x1}l_{x2})G(k_ynl_{x2}l'_x),\label{dyson}\\
\Sigma(k_ynl_xl'_x)&=&-\frac{T}{N}\sum_{k'_y\nu}\sum^1_{s,s'=-1} \theta(l_x+s)\theta(l'_x+s')\nonumber \\
&&\times G(k_y-k'_y,n-\nu,l_x+s,l'_x+s')\nonumber\\
&&\times\left[g_{k_y-k'_y,s}g_{k_ys'}D_{12}(k'_y\nu l_x,l'_x+s') \right.\nonumber\\
&&\quad+g_{k_y-k'_y,s}g_{k_y-k'_y,s'}D_{11}(k'_y\nu l_xl'_x) \nonumber\\
&&\quad+g_{k_ys}g_{k_ys'}D_{22}(k'_y\nu,l_x+s,l'_x+s') \nonumber\\
&&\quad\left.+g_{k_ys}g_{k_y-k'_y,s'}D_{21}(k'_y\nu,l_x+s,l'_x)\right], \nonumber\\
&&\label{sei}
\end{eqnarray}
where $n$ and $\nu$ are shorthand symbols for the Matsubara frequencies $\omega_n=(2n+1)\pi T$ and $\omega_\nu=2\nu\pi T$, respectively, $T$ is the temperature,
$$G^{(0)}(nl_xl'_x)=\delta_{l_xl'_x}\left(i\omega_n-\varepsilon_{l_x} \right)^{-1},$$
$\varepsilon_{l_x}=\mu-\frac{J}{2}\delta_{l_x0}$ with the last term taking into account the attraction of a hole to the boundary, $N$ is the number of sites in the $y$ direction,
$$g_{k_ys}=\left\{\begin{array}{ll}
                           2t\cos(k_y), & s=0, \\
                           t, & s=\pm 1, \\
                         \end{array}
                   \right. $$
and $D_{ij}(k_y\nu l_xl'_x)$ is the Fourier transforms of the components of the matrix magnon Green's function
\begin{eqnarray}
\hat D(k_y\tau l_xl'_x)&=&-\left\langle{\cal T}\hat B_{k_yl_x}(\tau)\hat B^\dagger_{k_yl_x}\right\rangle,\nonumber\\[-1.5ex]
&&\label{md}\\[-1.5ex]
\hat B_{k_y l_x}&=&\left(
                         \begin{array}{c}
                           b_{k_y l_x} \\
                           b^\dag_{-k_y,l_x} \\
                         \end{array}
                   \right).\nonumber
\end{eqnarray}

Equation~(\ref{sei}) describes the contribution of the sunrise diagram. In the case of the semi-infinite crystal there is also a nonzero contribution of the bubble diagram, which vanishes in an unbounded crystal. One can show, however, that in the semi-infinite crystal in the case of small hole concentrations the contribution of the bubble diagram is also negligibly small. Indeed, this term contains the multiplier
\begin{equation}
\sum_{k_y}g_{k_ys}\langle h_{k_y,l_x+s}h^\dagger_{k_yl_x}\rangle,
\label{mult}\end{equation}
where the mean value can be expressed through the retarded hole Green's function $G(k_y\omega l_xl'_x)$ as
$$\langle h_{k_yl_x}h^\dagger_{k_yl'_x}\rangle=- \int^\infty_{-\infty}\frac{d\omega}{\pi}\frac{{\rm Im}\,G(k_y\omega l_xl'_x)}{1+e^{-\omega/T}}.$$
For $T=0$ the integration is carried out over unoccupied states. For small hole concentrations these are in fact all states and therefore one can rewrite the above integral as
$$-\int^\infty_{-\infty}\frac{d\omega}{\pi}{\rm Im}\,G(k_y\omega l_xl'_x)=\delta_{l_xl'_x}.$$
Thus, for $s=\pm 1$ the multiplier~(\ref{mult}) is small because the mean value is negligible, while for $s=0$ it is small because $g_{k_y0}\propto\cos(k_y)$ and the sum over $k_y$ is negligible.

Let us switch from the Matsubara Green's functions to the real-frequency retarded Green's functions. It can be done using the following relation between these functions
\begin{widetext}
\begin{equation}
D_{ij}(k_y\nu l_xl'_x)=\int^\infty_{-\infty}\frac{d\omega}{2\pi} \frac{{\rm Im}\left[D_{ij}(k_y\omega l_xl'_x)+D_{ji}(k_y\omega l'_xl_x)\right]-i{\rm Re}\left[D_{ij}(k_y\omega l_xl'_x)- D_{ji}(k_y\omega l'_xl_x)\right]}{\omega-i\omega_\nu}.
\label{matsret}\end{equation}
The relation can be verified using the spectral representations. An analogous relation can be written for the hole Green's functions. From Eqs.~(\ref{dyson}), (\ref{sei}) and equations given below one can see that
$$
D_{ij}(k_y\omega l_xl'_x)=D_{ji}(k_y\omega l'_xl_x), \quad
G(k_y\omega l_xl'_x)=G(k_y\omega l'_xl_x).
$$
Thus, only imaginary parts of the retarded Green's functions appear in Eq.~(\ref{matsret}). Substituting these representations into self-energy~(\ref{sei}) and carrying out the summation over $\nu$ we find
\begin{eqnarray}
{\rm Im}\Sigma(k_y\omega l_xl'_x)&=&-\frac{1}{N}\sum_{k'_y} \sum^1_{s,s'=-1}\theta(l_x+s)\theta(l'_x+s')\int^\infty_{-\infty} \frac{d\omega'}{\pi}{\rm Im}G(k_y-k'_y,\omega-\omega',l_x+s,l'_x+s') \nonumber\\
&&\times\left[n_F(\omega'-\omega)+n_B(\omega')\right]\nonumber\\
&&\times\left[g_{k_y-k'_y,s}g_{k_ys'}{\rm Im}D_{12}(k'_y\omega' l_x,l'_x+s') +g_{k_y-k'_y,s}g_{k_y-k'_y,s'}{\rm Im}D_{11}(k'_y\omega' l_xl'_x)\right.\nonumber\\
&&\quad\left.+g_{k_ys}g_{k_ys'}{\rm Im}D_{22}(k'_y\omega',l_x+s,l'_x+s')
+g_{k_ys}g_{k_y-k'_y,s'}{\rm Im}D_{21}(k'_y\omega',l_x+s,l'_x)\right],
\label{se}
\end{eqnarray}
\end{widetext}
with $n_F(\omega)=\left(e^{\omega/T}+1\right)^{-1}$ and $n_B(\omega)=\left(e^{\omega/T}-1\right)^{-1}$. The real part of self-energy~(\ref{se}) can be calculated from the Kramers-Kronig relation. Self-energy equation~(\ref{dyson}) is transformed to real frequencies by the substitution $i\omega_n \rightarrow\omega+i\eta$, $\eta\rightarrow+0$.

In the considered case of small hole concentrations we can neglect the influence of holes on magnon Green's function~(\ref{md}) and use its value for the undoped case described by Hamiltonian~(\ref{haf}). In this case Green's function reads \cite{voropajeva,sherman09}
\begin{eqnarray}
\hat D(k_y\omega l_xl'_x)&=&\hat D^{(0)}(k_y\omega l_xl'_x) -\frac{J}{2}\hat D^{(0)}(k_y\omega l_x0)\nonumber\\
&&\times\left[\hat I+\frac{J}{2} \hat D^{(0)}(k_y\omega 00)\right]^{-1}\hat D^{(0)}(k_y\omega 0l'_x),\nonumber\\
&&\label{gfd}
\end{eqnarray}
where $\hat I$ is a $2\times 2$ identity matrix,
\begin{eqnarray}
&&\hat{D}^{(0)}(k_y\omega l_xl'_x)=\int_0^\pi dk_x\sin[k_x(l_x+1)] \sin[k_x(l'_x+1)]\nonumber\\
&&\quad\quad\times\left(\frac{\hat{P}_{\bf k }}{\omega-E_{\bf k}+i\eta} -\frac{\hat{Q}_{\bf k}}{\omega+E_{\bf k}+i\eta}\right),
\label{gfd0}\\
&&\hat{P}_{\bf k}=\left(
 \begin{array}{cc}
 u^2_{\bf k} & u_{\bf k}v_{\bf k} \\
 u_{\bf k}v_{\bf k} & v^2_{\bf k} \\
 \end{array}
\right),\quad
\hat{Q}_{\bf k}=\left(
 \begin{array}{cc}
 v^2_{\bf k} & u_{\bf k}v_{\bf k} \\
 u_{\bf k}v_{\bf k} & u^2_{\bf k} \\
 \end{array}
\right),\nonumber
\end{eqnarray}
${\bf k}=(k_x,k_y)$, $E_{\bf k}=2J\sqrt{1-\gamma^2_{\bf k}}$ is the bulk magnon energy, $\gamma_{\bf k}=\frac{1}{2}\left[\cos(k_x)+ \cos(k_y)\right]$, and
\begin{eqnarray*}
u_{\bf k}&=&\frac{1}{2}\left(\sqrt[4]{\frac{1-\gamma_{\bf k}}{1+\gamma_{\bf k}}}+\sqrt[4]{\frac{1+\gamma_{\bf k}}{1-\gamma_{\bf k}}}\right),\\
v_{\bf k}&=&\frac{1}{2}\left(\sqrt[4]{\frac{1-\gamma_{\bf k}}{1+\gamma_{\bf k}}}-\sqrt[4]{\frac{1+\gamma_{\bf k}}{1-\gamma_{\bf k}}}\right).
\end{eqnarray*}
In Eq.~(\ref{gfd}), the first term in the right-hand side describes the bulk modes -- the standing spin waves, while the second term is connected with the boundary spin waves. Their peak dominates in the spectral intensity $-{\rm Im}D(k_y\omega l_xl_x)$ for $l_x=0,1$ and practically disappears in site rows more distant from the boundary. \cite{voropajeva,sherman09}

It is instructive to elucidate how the equations obtained above are transformed to the form for an unbounded crystal with distance from the boundary. In Eq.~(\ref{gfd}), the second term in the right-hand side becomes negligibly small {\em if at least one of the coordinates} $l_x$ or $l'_x$ is larger than 2. Green's function $\hat{D}^{(0)}(k_y\omega l_xl'_x)$, to which $\hat{D}(k_y\omega l_xl'_x)$ is reduced for such $x$ coordinates, contains the multiplier $\sin[k_x(l_x+1)] \sin[k_x(l'_x+1)]$ in its integrand [see Eq.~(\ref{gfd0})]. If in this multiplier the sines are replaced by their representation through exponential functions, one can realize that terms with the same signs of exponents are small for large $l_x$ or $l'_x$, since the respective exponential functions rapidly oscillate. Remaining terms depend only on the difference $l_x-l'_x$ as it must for the unbounded crystal. It can be shown that these terms are identical to Green's function for this case. Since the magnon Green's function defines the form of the hole Green's function, one can expect that the latter also becomes close to its unbounded form when at least one of the $x$ coordinates is large. Taking this into account, from Eqs.~(\ref{dyson}) and (\ref{se})-(\ref{gfd0}) after the Fourier transformation we obtain equations for the unbounded crystal of Refs.~\onlinecite{marsiglio,martinez}.

The above discussion allows us to transform Eq.~(\ref{dyson}) into a more tractable form. Let us rewrite it as
\begin{eqnarray}
&&\sum^{l_{xm}}_{l''_x=0}\left[\left(\omega-\varepsilon_{l_x}\right) \delta_{l_xl''_x}-\Sigma(k_y\omega l_xl''_x)\right]G(k_y\omega l''_xl'_x)\nonumber\\
&&\quad\quad=\delta_{l_xl'_x}+M(k_y\omega l_xl'_x),\label{dyson2}
\end{eqnarray}
where
\begin{equation}\label{mmatrix}
M(k_y\omega l_xl'_x)=\sum_{l''_x>l_{xm}}\Sigma(k_y\omega l_xl''_x)G(k_y\omega l''_xl'_x).
\end{equation}
In Eq.~(\ref{dyson2}), we assume that the coordinates $l_x$ and $l'_x$ are restricted within the range $[0,l_{xm}]$. The parameter $l_{xm}$ is expected to be large enough for substituting the self-energy and Green's function in Eq.~(\ref{mmatrix}) by their values in an unbounded crystal, in compliance with the above discussion. At the same time $l_{xm}$ can be chosen to be small enough for the inversion of the matrix in the left-hand side of Eq.~(\ref{dyson2}) would not lead to time-consuming calculations.

\section{Results and discussion}
In the below calculations we set $T=0$ and $J/t=0.2$. The latter ratio of parameters corresponds to hole-doped cuprates. \cite{mcmahan,gavrichkov} Equations~(\ref{dyson2}) and~(\ref{mmatrix}) were solved by iterations for $l_{xm}=4$, using as the starting function for $G(k_y\omega l_xl'_x)$ Green's function of an unbounded crystal. To ensure the convergence of the iteration procedure an artificial broadening was introduced by substituting $\omega$ with $\omega+i\eta$, $\eta=0.05t$, in Eq.~(\ref{dyson2}). The chemical potential $\mu$ was chosen so that the frequency $\omega=0$, which separates occupied and unoccupied states, was located in the low-frequency tail of the spectral function
$$A(k_y\omega l_x)=-{\rm Im}G(k_y\omega l_xl_x).$$
This ensures a low hole concentration expected in the derivation of the above formulas. \cite{remark}

\begin{figure}
\centerline{\includegraphics[width=8cm]{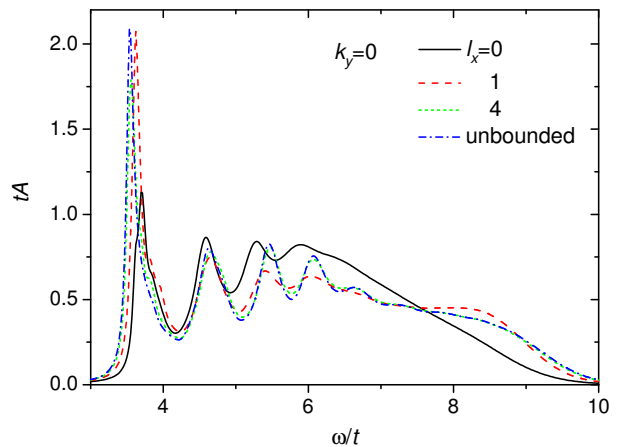}}
\caption{(Color online) The spectral function $A(k_y\omega l_x)$ for $k_y=0$, $l_x=0$, 1, 4 and in an unbounded crystal.} \label{Fig1}
\end{figure}

The spectral function gives the density of states projected on states of the row $l_x$. A typical example of this quantity obtained in the course of the calculations is shown in Fig.~\ref{Fig1}.
Besides the spectral function for near-boundary rows, Fig.~\ref{Fig1} contains also the spectral function of the unbounded crystal,
\begin{eqnarray*}
A_b(k_y\omega,l_x-l'_x)&=&-\frac{1}{2\pi}\int^\pi_{-\pi}dk_x \cos\left[k_x\left(l_x-l'_x\right)\right]\\
&&\quad\quad\quad\times{\rm Im}G_b(k_yk_x\omega),
\end{eqnarray*}
which is given for comparison. Due to the translation symmetry this function depends only on the difference $l_x-l'_x$ and for the considered case $l_x=l'_x$ its last argument is zero. In shape this function resembles spectral functions obtained for a fixed wave vector $k_x$ in an unbounded crystal. \cite{marsiglio,martinez}  However, the maxima in Fig.~\ref{Fig1} are somewhat broadened in comparison with these functions due to the integration over $k_x$ in the above formula. As would be expected, the spectrum in the boundary row $l_x=0$ differs most greatly from $A_b(k_y\omega 0)$. From the figure one can see how the spectrum is transformed, gradually approaching to the spectrum of an unbounded crystal, with distance from the boundary. In the scale of Fig.~\ref{Fig1} already the spectrum in the fifth row ($l_x=4$) is barely distinguishable from $A_b(k_y\omega 0)$.

\begin{figure}
\centerline{\includegraphics[width=8cm]{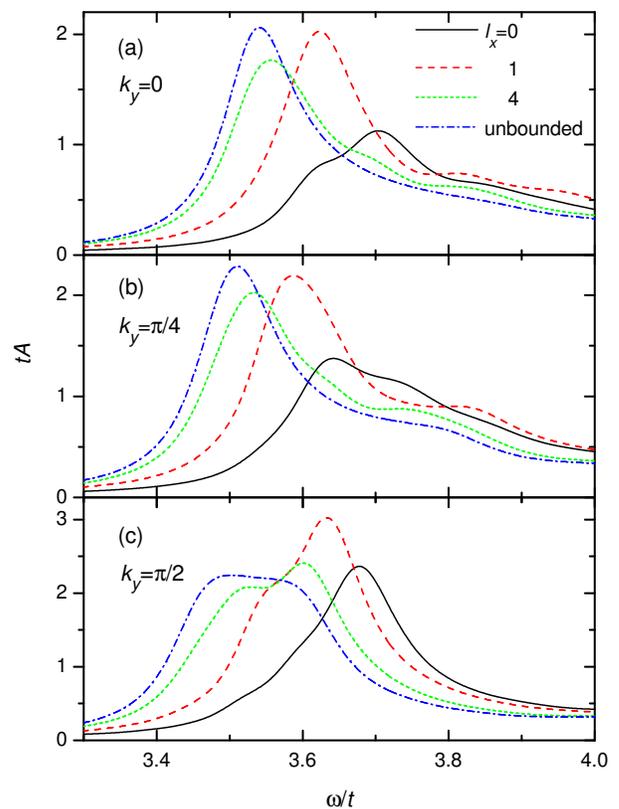}}
\caption{(Color online) The spectral function $A(k_y\omega l_x)$ in the vicinity of the main maximum for $k_y=0$ (a), $\pi/4$ (b) and $\pi/2$ (c) in the rows $l_x=0$, 1, 4 and in an unbounded crystal.} \label{Fig2}
\end{figure}
The vicinity of the main maximum of the spectral function is shown in Fig.~\ref{Fig2} for several wave vectors. From formulas of the previous section it can be shown that $A(k_y\omega l_x)=A(\pi-k_y,\omega l_x)$. Besides, in accord with the symmetry $A(k_y\omega l_x)=A(-k_y,\omega l_x)$. From these figure and equations one can see that the spectral maxima are shifted to higher frequencies on approaching the boundary for all wave vectors $k_y$. In accordance with this the low-frequency tails of the maxima become weaker with decreasing $l_x$. Since the concentration of holes in a row $x(l_x)$ is determined by this tail,
\begin{eqnarray*}
x(l_x)&=&\frac{1}{N}\sum_{k_y}\left\langle h^\dagger_{k_yl_x}h_{k_yl_x} \right\rangle\\
&=&\frac{1}{N}\sum_{k_y}\int^\infty_{-\infty} \frac{d\omega}{\pi}A(k_y\omega l_x)n_F(\omega),
\end{eqnarray*}
one can conclude that the concentration decreases monotonically on approaching the boundary -- {\em near-boundary rows are depleted of holes}.

To elucidate a formation mechanism of this hole depletion layer let us first consider the role of two above-mentioned factors, which can influence the population of holes in the near-boundary region -- the attraction of a hole to the boundary and the near-boundary magnon mode. For the chosen chemical potential, the energy of an immobile hole is equal to $6t$. As seen in Figs.~\ref{Fig1} and~\ref{Fig2}, by virtue of the interactions there is the energy gain equal approximately to $2.5t$ in states corresponding to the main maximum. Therefore, the attraction which is of the order of $J\ll t$ plays practically no role in the hole distribution. This conclusion is confirmed by calculations -- omitting the attraction term is barely perceptible in the shape and location of maxima of the spectral function.
\begin{figure}
\centerline{\includegraphics[width=8cm]{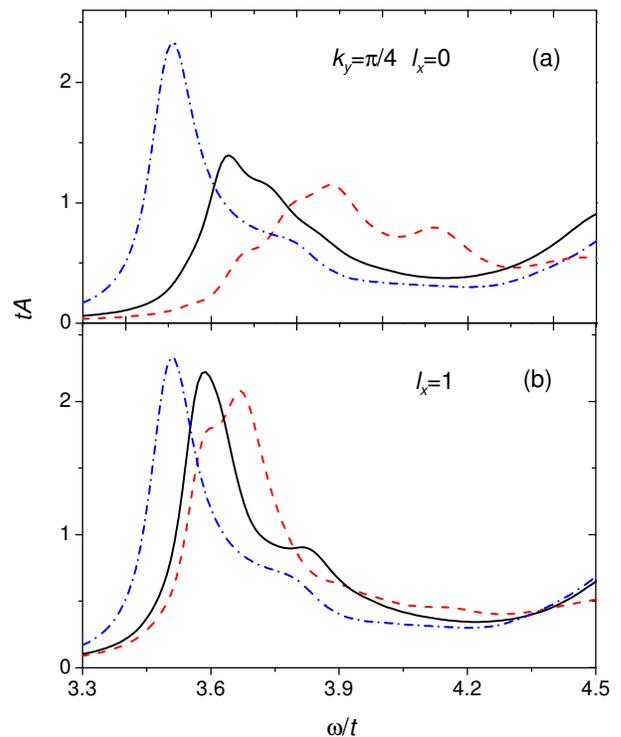}}
\caption{(Color online) The spectral function $A(k_y\omega l_x)$ in the vicinity of the main maximum with taking into account the boundary magnon mode (black solid lines) and without it (red dashed lines) for $k_y=\pi/4$, $l_x=0$ (a) and 1 (b). Blue dash-dotted lines corresponds to $A_b(k_y\omega 0)$.} \label{Fig3}
\end{figure}
The contribution of the near-boundary magnon mode can be evaluated from Fig.~\ref{Fig3}. The spectral function without this mode was calculated with the magnon Green's function~(\ref{gfd0}) instead of the full function~(\ref{gfd}). As seen from the figure, the near-boundary mode makes its contribution in the location of the maximum and the intensity redistribution. This contribution is especially detectable for the boundary row. However, with this mode or without it the main maxima in the near-boundary rows have higher frequencies than deep inside the crystal. Consequently, the near-boundary mode does not play the main role in the formation of the hole depletion layer.

To understand the appearance of the depletion layer let us remind that in the considered model holes are spin polarons. \cite{marsiglio,martinez} Due to the antiferromagnetic background a hole can move over the lattice only with the emission and absorption of magnons, as it is seen from Hamiltonian~(\ref{hsw}). As a consequence the hole is surrounded by a cloud of magnons. Without spins the maximum energy gain which a moving hole can achieve in comparison with an immobile quasiparticle is $4t$ -- the difference between the lowest energy in the 2D nearest-neighbor band and its center of mass. In the antiferromagnetic lattice this gain is decreased by the energy consumption for the distortion of the magnetic order around the hole. For the ratio $J/t=0.2$ the energy gain is reduced approximately to $2.5t$ (see the above figures). This energy gain is still comparable with the maximal possible value $4t$. Notice that at the same time the spin polaron bandwidth is of the order of $J$ for low doping, which is much smaller in comparison with the energy gain. \cite{marsiglio,martinez} This large energy gain complicates the formation of ferrons -- ferromagnetically ordered regions around holes -- and stripes in the $t$-$J$ model. Only for very small ratios $J/t$ the gain in the hole kinetic energy in the ferromagnetic region becomes large enough to stabilize ferrons. \cite{nagaev,hizhnyakov,sherman90} Away from the boundary the magnon cloud has the symmetry determined by the group of the hole wave vector and this symmetry ensures the lowest energy of the spin polaron. Near the boundary, the cloud is distorted, which lowers the symmetry and inevitably leads to a growth of the energy. It is the mechanism of the depletion layer formation in the considered model. The depth of the row in which the location of the main maximum coincides with that in an unbounded crystal gives an estimate of the magnon cloud size. In our case, its radius is equal to 4 lattice spacings.

Notice that as in Refs.~\onlinecite{hasegawa,potthoff97,potthoff99,ishida} in our case the decrease in the spectral intensity of the main maximum in the boundary row is connected with the reduced boundary coordination number. In the mentioned works this leads to an effective strengthening of the Hubbard repulsion on the boundary, while in our case to the deformation of the magnon cloud around the hole in the spin polaron.

\begin{figure}
\centerline{\includegraphics[width=8cm]{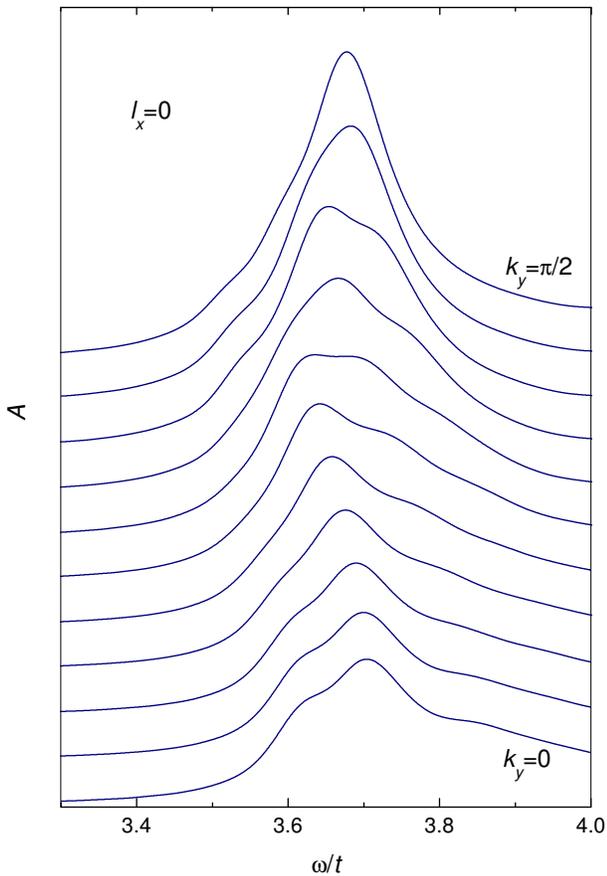}}
\caption{(Color online) The spectral function $A(k_y\omega l_x)$ in the vicinity of the main maximum for $l_x=0$ and $k_y$ ranging from 0 (the bottom curve) to $\pi/2$ (the upper curve) with the step $\pi/20$. For better visibility curves with larger $k_y$ are shifted upward with respect to curves with smaller wave vectors.} \label{Fig4}
\end{figure}
\begin{figure}
\centerline{\includegraphics[width=8cm]{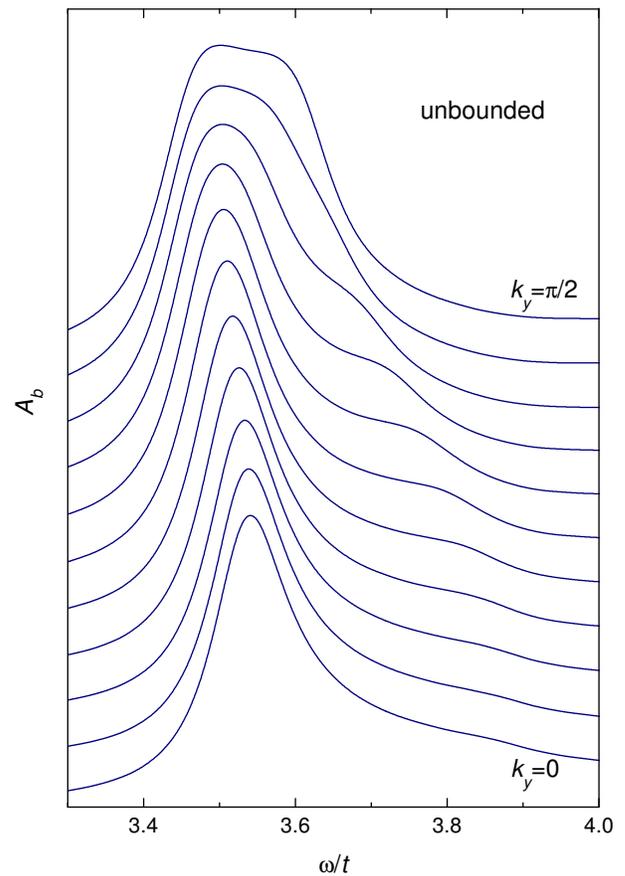}}
\caption{(Color online) The same as Fig.~\protect\ref{Fig4}, but for the spectral function of an unbounded crystal $A_b(k_y\omega 0)$.} \label{Fig5}
\end{figure}
Closer inspection of the obtained spectral functions shows that the main maximum for the boundary row has a more complicated structure than spectra for $l_x=2$ to 4 and for the unbounded crystal. Figures~\ref{Fig4} and~\ref{Fig5} demonstrate this difference. The evolution of maxima in rows $l_x=2$ to 4 are similar to that shown in Fig.~\ref{Fig5}, while for $l_x=1$ the spectrum has some features of the boundary row. This result demonstrates how the deeper near-boundary regions approach in their properties to the unbounded crystal, which states are characterized by a 2D wave vector. As known, \cite{marsiglio,martinez} the energy of these states has a minimum at the points $\left(\pm\frac{\pi}{2},\pm\frac{\pi}{2}\right)$ and their dispersion is weak on the boundary of the magnetic Brillouin zone, which is composed of segments $(0,\pm\pi)-(\pm\pi,0)$. The states near these segments make the main contribution into the maximum in Fig.~\ref{Fig5} -- for $k_y=0$ wave vectors of these states lie near $(\pm\pi,0)$, while for $k_y=\frac{\pi}{2}$ these wave vectors are from the neighborhood of $\left(\pm\frac{\pi}{2},\frac{\pi}{2}\right)$. The change in the location of the maximum when $k_y$ grows from 0 to $\frac{\pi}{2}$ in Fig.~\ref{Fig5} reflects the mentioned weak dispersion of the states along the boundary of the magnetic Brillouin zone. The shoulder, which approaches the maximum from high frequencies, is mainly connected with states from the vicinity of the axes and the boundary of the Brillouin zone -- for $k_y=\frac{\pi}{2}$ these states have wave vectors near $(0,\frac{\pi}{2})$ and $(\pm\pi,\frac{\pi}{2})$. A similar high-frequency shoulder is observed in Fig.~\ref{Fig4}. However, in contrast to the deeper rows, the main maximum in the boundary row has also a low-frequency shoulder which is best seen for small $k_y$. As follows from Fig.~\ref{Fig2}(a), the locations of this latter shoulder is close to the position of the maximum in the row $l_x=1$. Indeed, in the considered system two neighboring rows have maxima, which are shifted in the frequency scale relative to each other. Since in accord with the formulas of the previous section the spectral function of a given row is connected with functions in neighboring rows, one can expect that a replica of the more intensive maximum for $l_x=1$ will be seen in the boundary row. In the present case this replica looks like the low-frequency shoulder of the main maximum. Thus, a more complicated character of the boundary spectra is connected with the replica of the maximum of the underlying row. Notice that this replica is an attendant effect of the hole depletion in the near-boundary region.

\section{Conclusion}
Our calculations referred to a 2D crystal. From the similarity of the 2D and 3D magnon spectra \cite{voropajeva,sherman09} we can expect to obtain analogous results for charge carriers in a 3D crystal with strong electron correlations. From these results, the conclusion can be drawn that the surface electronic structure, which is tested by the photoelectron and tunnel spectroscopies, even in the considered case of the idealized surface may essentially differ from the bulk spectrum. The discrepancies between the photoemission data of a number of transition-metal oxides and calculated bulk spectra were interpreted similarly in Refs.~\onlinecite{matzdorf,maiti}.

Comparing results obtained in the semi-infinite Hub\-bard \cite{hasegawa,potthoff97,potthoff99,ishida} and $t$-$J$ models, we find some similar features. In spite of the differences of models and computation methods, in both models for uniform parameters the quasiparticle weight is lowered, while the intensity of the high-energy part of the spectrum grows \cite{potthoff97} at the boundary. The reason for this intensity redistribution is similar -- it is a reduced coordination number at the boundary, which leads to an effective strengthening of the on-site repulsion in the Hubbard model and to the deformation of the magnon cloud around the hole in the spin polaron in the $t$-$J$ model.

In summary, in the present article, we investigated the spectral function of the 2D $t$-$J$ model on a semi-infinite lattice. The limit of strong electron correlations, $t\gg J$, and the case of low hole concentrations were considered. For this investigation, we used the spin-wave approximation and the diagram technique with the non-crossing approximation. The obtained self-energy equations were solved by iterations, and we could trace the variation of the spectral function with distance from the boundary. Already in the fifth row the spectral function nearly coincided with its bulk counterpart. It was shown that the near-boundary region of the crystal is depleted of holes. The reason is the deformation of a magnon cloud around a hole in this region, which is accompanied by energy losses. The hole depletion is reflected in dissimilar locations and intensities of the main spectral maxima for different site rows near the boundary. As a consequence a replica of a maximum in the second row is seen in the boundary spectral function. This results in its more complicated shape in comparison with the bulk spectrum.

\begin{acknowledgments}
This work was supported by the ETF grant No.~6918.
\end{acknowledgments}


\begin{thebibliography}{99}
\bibitem{dagotto}E.~Dagotto, Science {\bf 318}, 1076 (2007).
\bibitem{hasegawa}H.~Hasegawa, J.\ Phys.: Condens.\ Matter {\bf 4}, 1047 (1992).
\bibitem{potthoff97}M.~Potthoff and W.~Nolting, Z.\ Phys.\ B {\bf 104}, 265 (1997).
\bibitem{potthoff99}M.~Potthoff and W.~Nolting, Phys.\ Rev.\ B {\bf 59}, 2549 (1999); {\bf 60}, 7834 (1999).
\bibitem{ishida}H.~Ishida and A.~Liebsch, Phys.\ Rev.\ B {\bf 79}, 045130 (2009).
\bibitem{hirsch85}J.~E.~Hirsch, Phys.\ Rev.\ B {\bf 31}, 4403 (1985).
\bibitem{dagotto94}E.~Dagotto, Revs.\ Mod.\ Phys.\ {\bf 66}, 763 (1994).
\bibitem{hoglund}K.~H.~H\"{o}glund and A.~W.~Sandvik, Phys.\ Rev.\ B {\bf 79}, 020405(R) (2009).
\bibitem{metlitski}M.~A.~Metlitski and S.~Sachdev, Phys.\ Rev.\ B {\bf 78}, 174410 (2008).
\bibitem{pardini}T.~Pardini and R.~R.~P.~Singh, Phys.\ Rev. B {\bf 79} 094413 (2009).
\bibitem{voropajeva}N.~Voropajeva and A.~Sherman, arXiv:0912.4958  (unpublished).
\bibitem{sherman09}N.~Voropajeva and A.~Sherman, Phys.\ Lett.\ A  {\bf 373}, 3473 (2009); A.~Sherman and N.~Voropajeva, Intern.\ J.\ Modern Phys.\ B {\bf 24}, 979 (2010).
\bibitem{marsiglio}F.~Marsiglio, A.~E.~Ruckenstein, S.~Schmitt-Rink, and C.~M.~Varma, Phys.\ Rev.\ B {\bf 43}, 10882 (1991).
\bibitem{martinez}G.~Martinez and P.~Horsch, Phys.\ Rev.\ B {\bf 44}, 317 (1991).
\bibitem{hirsch}J.~E.~Hirsch, Phys.\ Rev.\ Lett.\ {\bf 59}, 228 (1987).
\bibitem{izyumov}Yu.~A.~Izyumov and Yu.~N.~Skryabin, {\em Statistical Mechanics of Magnetically Ordered Systems} (Consultants Bureau, New York, 1988).
\bibitem{tyablikov}S.~V.~Tyablikov, {\em Methods of the Quantum Theory of Magnetism} (Plenum Press, New York, 1967).
\bibitem{mcmahan}A.~K.~McMahan, J.~F.~Annett, and R.~M.~Martin, Phys.\ Rev.\ B {\bf 42}, 6268 (1990).
\bibitem{gavrichkov}V.~A.~Gavrichkov, S.~G.~Ovchinnikov, A.~A.~Borisov, and E.~G.~Goryachev, Zh.\ Eksp.\ Teor.\ Fiz.\ {\bf 118}, 422 (2000) [JETP (Russia) {\bf 91}, 369 (2000)].
\bibitem{remark}It is worth noting that the shape of the spectral function is markedly changed only when the frequency of the main maximum $\omega_m$ becomes close to $\omega=0$, see, e.g., A.~Sherman and M.~Schreiber, Phys.\ Rev.\ B {\bf 50}, 12887 (1994). Thus, nearly the same spectral function as shown in Fig.~\ref{Fig1} is obtained for any chemical potential for which $\omega_m\agt t$.
\bibitem{nagaev}E.~L.~Nagaev, Phys.\ Rev.\ B {\bf 64}, 014401 (2001).
\bibitem{hizhnyakov}V.~Hizhnyakov and E.~Sigmund, Physica C {\bf 156}, 655 (1988).
\bibitem{sherman90}A.~Sherman, Physica C {\bf 171}, 395 (1990); J.~Sabczynski, M.~Schreiber, and A.~Sherman, Phys.\ Rev.\ B {\bf 48}, 543 (1993).
\bibitem{matzdorf}R.~Matzdorf, Z.~Fang, Ismail, J.~Zhang, T.~Kimura, Y.~Tokura, K.~Terakura, and E.~W.~Plummer, Science {\bf 289}, 746 (2000).
\bibitem{maiti}K.~Maiti, D.~D.~Sarma, M.~J.~Rozenberg, I.~H.~Inoue, H.~Makino, O.~Goto, M.~Pedio, and R.~Cimino, Europhys.\ Lett.\ {\bf 55}, 246 (2001).
\end{thebibliography}
\end{document}